\documentclass[twoside,preprintnumbers,superscriptaddress,amsmath,amssymb,11pt,pacs,nofootinbib]{revtex4}
\usepackage{amsmath,amssymb,url}
\usepackage{graphicx,pslatex,subfigure}
\usepackage{esint}
\usepackage{lscape}
\usepackage{pstricks}
\usepackage{color}
\usepackage{braket,euscript}
\usepackage{fancyvrb}
 \usepackage{hyperref}

\begin{document}

\title{{\bf  About the triviality of the higher derivative sector in the Abelian Lee-Wick model}}
\author{D.~Fiorentini}
\email{diegofiorentinia@gmail.com}
\affiliation{Departamento de F\'{\i}sica Te\'{o}rica,
Instituto de F\'{\i}sica, UERJ - Universidade do Estado do Rio de Janeiro, Rua S\~ao Francisco Xavier 524, 20550-013, Maracan\~a, Rio de Janeiro, Brasil}
\author{V.~J.~Vasquez~Otoya}
\email{victor.vasquez@ifsudestemg.edu.br}
\affiliation{IFSEMG $-$ Instituto Federal de Educac\~ao, Ci\^encia e Tecnologia, Rua Bernardo Mascarenhas 1283, 36080-001, Juiz de Fora, MG, Brasil}

\begin{abstract}

Canonical quantization is reviewed here for the Abelian Lee-Wick model by using the Dirac constraints method, a Gupta-Bleuler-like prescription is implemented and the BRST charge operator for this model is built. New degrees of freedom associated to the (higher derivatives) Lee-Wick ghost are excluded from the (physical) space of observables by the imposition of a Piguet-Nielsen-type BRST-extended symmetry for the Lee-Wick mass term, removing the Lee-Wick mass parameter of the theory, which implies remove all the higher-derivative sector. In addition, another prescription based on a new subsidiary condition of invariance under a purely higher-derivative BRST-type new symmetry is proposed.

\end{abstract}
\maketitle

\section{Introduction to Lee-Wick Model}

T.D. Lee and G.C. Wick (LW) proposed \cite{Lee:1969fy,Lee:1970iw} a finite ultraviolet version for the Quantum Electrodynamics (QED) by associating Pauli-Villars cutoff \cite{Pauli:1949zm} with an additional (fundamental) massive ghost field. To do that, one adds up a higher derivative (HD) kinetic term with wrong-sign with respect to the usual Abelian theory. This additional degrees of freedom regularize the theory if they correspond to an indefinite metric in the Hilbert space\footnote{The alternative is to define these new particles as negative energy states, however, in so doing, the unitarity is lost.} with the LW modes removed from the physical space by directly constraining, guaranteeing the unitarity of the $S-$matrix \cite{Lee:1970iw,Cutkosky:1969fq,Lee:1971ix,Antoniadis:1986tu}. The regulator modes (negative norm states) would have a very large mass and width such that they would decay into usual lighter positive norm particles. However, there is a price to be paid: the theory presents acausal effects, although they do not appear to conflict with any experimental observations since the time scale (defined by the LW-mass) of this acausality is far too small \cite{COLEMAN1970282,vanTonder:2008ub,Grinstein:2008bg}. However, for high enough energy scale acausal effects arise, \textit{e.g.}, in a early universe stage. Another problem emerges if multi-ghosts are taken into account, because if they have a positive energy, the imposition of LW's constraint does not work. A mechanism to circumvent this problem was proposed by Lee and generalized by Cutkosky \textit{et al.} \cite{Cutkosky:1969fq} via boundary future conditions and the deformation of integration contours in the Feynman diagram order by order; however, an all-order perturbative proof of this mechanism is still unknown. Furthermore, these prescriptions are non-perturbatively meaningless except when Lorentz invariance is sacrificed \cite{Nakanishi:1971jj,Nakanishi:1971ky,Gleeson:1972xj}. 

With the modern understanding of renormalization of the Yang-Mills theories and with the general prescription to incorporate the Abelian theory in a unique unified framework, the original motivation of the LW model was lost. However, gravity, which is a non-renormalizable theory  \cite{'tHooft:1974bx,Deser:1974cz,Deser:1974xq}, cannot be incorporated into this unification program. Here, the LW ideas play a natural role: following the proposal by Utiyama and DeWitt \cite{Utiyama:1962sn} of higher-derivative terms by the inclusion of quadratic curvature terms into the Einstein-Hilbert action, Stelle \cite{Stelle:1976gc} proved that one gets a renormalizable theory. Boulware \cite{Boulware:1983yj} got that the LW prescription defines a consistent and stable theory. In the context of the Yang-Mills coupling to the non-renormalizable effective Einstein-Hilbert theory, it has been shown by Ebert, Plefka, and Rodigast \cite{Ebert:2007gf} that 1-loop divergences can be removed by adding a counter-term exactly given by the LW-term.

Recently, in the context of the gauge theories, LW-theories have been reborn in the form of the Lee-Wick Standard Model (LWSM), proposed by Grinstein, O{'}Connell and Wise \cite{Grinstein:2007mp} as an alternative to solve the hierarchy problem: nonsupersymmetric renormalizable theories with elementary scalar fields (\textit{e.g.}, Higgs fields) present quadratically divergent radiative corrections for this bosonic mass. Technically, these divergences can be reabsorbed into the renormalization factors, but having a scalar mass much smaller than the scale of the theory requires an extremely fine tuning for this cancellation, which appears non-natural. In LWSM, higher derivative LW-type terms are added for each elementary particle, which is associated to a new spectrum of particles (LW-partners), like in supersymmetry, but with the same statistics.  Since the propagators of ordinary and LW particles differ by an overall sign, quadratic divergences cancel between pairs of diagrams. This fact has motivated an extensive study of the phenomenology of LW models \cite{Carone:2008bs,Espinosa:2007ny,Dulaney:2007dx,Grinstein:2008gq,Wise:2009mi,
Chivukula:2010nw,Espinosa:2011js,Grinstein:2008bg}, experimental and theoretical signatures \cite{Rizzo:2007ae,Alvarez:2008za,Underwood:2008cr,Figy:2011yu,Carone:2014kla}, higher-derivative extensions \cite{Carone:2008iw,Carone:2009it,Cho:2010hj} and a finite temperature investigation \cite{Fornal:2009xc,Bhattacharya:2011bb,Lebed:2013gua}.   

There exists a most important reason for the study of LW theories: LW higher derivative terms, as proved by \cite{Moayedi:2013nxa}, emerge as a natural Lagrangian density deformation for the electrodynamics (in 3+1 dimensions) if a non-zero minimal length exists. This is a consequence of the induced deformation in the Lorentz algebra for quantized spaces with just one temporal dimension \cite{Quesne:2006is} and, of course, this is consistent with the (extended) uncertainty principle and string and quantum loops theories, which  predict the existence of a minimal length of the order of the Planck length \cite{Hossenfelder:2012jw}.

The aim of this work is to propose two new more general prescriptions for cancelling possible asymptotic states in LW-partners without recourse to the old prescriptions. First, entirely removing the higher derivative sector of the observables via an extended BRST transformation \textit{\`a la Nielsen-Piguet} \cite{Nielsen:1975fs,Nielsen:1975ph,Piguet:1984js} for including the LW mass parameters in a BRST-doublet. Second, by implementing a subsidiary condition \textit{\`a la Ojima-Nakanishi} for observables \cite{Ojima:1978hy}, which is possible due to a new BRST-type symmetry for the LW-partners. 
Formally, standard canonical quantization procedure allow us to construct a unambiguous theory even without positive definite norm and energy sates, but this leads indefectibly to anti-causality in the higher-derivative sector, \textit{i.e.}, the LW-partners are defined as particles that travel backwards in time.
However, since physical states correspond to  the quotient of the kernel of the BRST charge with its range, we propose move on ghost states to the trivial sector of the cohomology of the BRST(-type) operator.
The most simple form is extend the BRST-transformation via a doublet to include the LW mass, so that observable quantities do not depend on this parameter. Thus, the theory shall not include the new degrees of freedom in its physical sector, once they are always \textit{tied up} to the LW mass, which is now in the trivial sector of the cohomology of the usual BRST operator. Since the mass parameter is in the trivial sector, then the resulting theory is (in a classical level) physically identical to the Maxwell theory. This enables us that, \textit{e.g.}, in the Slavnov regularization scheme \cite{Bakeyev:1996is}, include higher derivative gauge invariant terms, removing divergences without change the physical space. In the second prescription, we show that all fields associated to HD terms are confined to virtual sector by forming a quartet (\textit{i.e.}, two doublets) with respect to the new BRST-type symmetry, resulting in a theory which can depend on the LW-mass parameter.
In the reference \cite{Rivelles:2003jd} was showed that a special class of  higher-derivative Lagrangians theories have a BRST-type symmetry under which its Lagrangians can be written as a pure variation of a BRST-type operator, like in the so-called cohomological theories of topological type \cite{Rivelles:2003jd,Birmingham:1991ty}, which specify their cohomological triviality. We propose to extend this ideas for the non-trivial case of the standard Abelian theory when LW term is taking into account. 
Whilst in \cite{Rivelles:2003jd} the presence of an auxiliary set of fields with the same vectorial structure that the gauge field is imposes, allowing him to properly build the aforementioned symmetry with a quartet structure, the aim of the section \ref{1stpre}, where the \textit{usual} BRST charge operator is building, is to show that this fields emerge naturally via the quantization procedure, and represent independent degrees of freedom  related to the gauge fixing in the HD sector. For this reason, higher derivative terms in the auxiliary sector can be rewrite as the vectorial fields which define a new HD-BRST-type symmetry, as proposed in \cite{Rivelles:2003jd}. In our case, however, the classical Lagrangian after the (Dirac) quantization procedure cannot be written as a exact BRST variation, but as a closed BRST form. 
Note that both the first and second prescription not exhibit the problem of possible stables physical states with acausal particles in a high enough energy scale, asa when usual prescriptions are used, which is the main outcome of this work.

The outline of our work is as follows: in Section 1, we briefly review the canonical quantization for the Abelian LW model by adopting Dirac brackets. In Section 2, we come to the Gupta-Bleuler approach, built a BRST operator and we present our first prescription to eliminate the full sector of higher derivations. Section 3 is devoted to explicitly show that the higher derivative sector is cohomologically trivial. Finally, we display our Conclusions in Section 5.

\section{Canonical Quantization for the LW-Abelian Theory via Dirac brackets}
\label{CQDB}

In this section, the canonical quantization for the LW Abelian model is reviewed in order to clarify the count of degrees of freedom and explicitly showing how the structure of first class constraints leads us to implement a gauge fixing in the HD sector. This will be very important later, since the new independent auxiliary sector will be used to build the BRST-type symmetry on the section \ref{BRSTtriv}.

The starting action of the Lee-Wick Abelian gauge theory is given by \cite{Lee:1969fy,Lee:1970iw}
\begin{equation}
\mathcal{L}=-\frac{1}{4}F_{\mu\nu}F^{\mu\nu}
+\frac{1}{4M^2}(\partial_{\alpha}F_{\beta\gamma})(\partial^{\alpha}F^{\beta\gamma})
\label{LWaction}
\end{equation}
where the field strength tensor $F_{\mu\nu}$ is defined as usually, in terms of the potential vector. This action can be understood as a usual Abelian model with an extra pole in the two-point correlation function near the mass $M$, which corresponds to the LW-partner for the photon field. The additional particle in this dipole spectrum possesses an opposite (ghost) signature, which leads to a $1/p^4$ leading behaviour for the photon propagator in the ultraviolet region, regularizing naturally divergences associated with self-energies. Clearly, in the limit $M\rightarrow\infty$ we are back to the Maxwell theory.
The respective equations of motion for this (higher derivative) Lagrangian can be easily derived following the Ostrogradski procedure \cite{Ostro}\footnote{A pedagogical review may be found in \cite{Whittaker:104872}}:
\begin{equation}
\left(1+\frac{\partial^2}{M^2}\right)\partial_{\nu}F^{\nu\mu}=0
\end{equation}
or, in terms of the potential field
\begin{equation}
\left(1+\frac{\partial^2}{M^2}\right)\partial^2A^\mu
-\partial^{\mu}\left(1+\frac{\partial^2}{M^2}\right)\partial_{\nu}A^\nu=0
\end{equation}
 Notice that the kinetic matrix, i.e., the symmetric matrix that contains the coefficients of the quadratic terms  in \eqref{LWaction} is a polynomial of degree $2\times 2$. This means that we must double the dimension of the phase space, \textit{i.e.}; the Hamiltonian must be written as a function of the following two pairs of canonical variables \cite{Ostro}
\begin{eqnarray}
&Q_{1,\mu}=A_{\mu}\,,\qquad \Pi_{\mu}=\frac{\delta\mathcal{L}}{\delta A_{\mu}} \\
&Q_{2,\mu}=\partial_t A_{\mu}\equiv \tilde{A}_{\mu}\,,\qquad \tilde{\Pi}_{\mu}=\frac{\delta\mathcal{L}}{\delta \partial_t A_{\mu}}\equiv\frac{\delta\mathcal{L}}{\delta \tilde{A}_{\mu}}
\end{eqnarray}
provided that they satisfy the equal-time Poisson Brackets:
\begin{eqnarray}
\{A_{\mu}(x),\Pi^{\nu}(y)\}_{PB}&=&-\{\Pi^{\nu}(y),A_{\mu}(x)\}_{PB}\nonumber\\
\{\tilde{A}_{\mu}(x),\tilde{\Pi}^{\nu}(y)\}_{PB}&=&-\{\tilde{\Pi}^{\nu}(y),\tilde{A}_{\mu}(x)\}_{PB}\nonumber\\
&=&\delta_{\mu\nu}\delta^3(x-y)
\label{PB}
\end{eqnarray}
and all the rest equal to zero. By introducing the new field, $\tilde{A}_{\mu}$, we are converting the original higher derivative theory into an equivalent (effective) first order theory by adding a constraint which actually imposes that
$\tilde{A}_{\mu}$ is the time derivative of $A_{\mu}$ via a temporal dependent Lagrange multiplier. In terms of the original Ostrogradski theory, LW-partners correspond to the so-called \textit{Ostrogradski ghost}. Then, the set canonically conjugate momenta is given by
\begin{eqnarray}
&\Pi^{\mu}=-F^{0\mu}+\frac{1}{M^2}\left(\partial^2F^{0\mu}+\partial_i\partial_0F^{i\mu}\right)\\
&\tilde{\Pi}^{\mu}=\frac{1}{M^2}\partial_0F^{0\mu}
\end{eqnarray}
so that, the canonical Hamiltonian is a function of four 4-components canonical variables and is given by
\begin{eqnarray}
H_c&=&H_c(A_{\mu},\Pi_{\mu},\tilde{A}_{\mu},\tilde{\Pi}_{\mu})\nonumber\\
&=&\Pi^{\mu}\partial_0A_{\mu}-\tilde{\Pi}^i\partial_0\tilde{A}_{\mu}-\mathcal{L}\nonumber\\
&=&\frac{1}{2}\left(E^iE_i+B^iB_i\right)+\frac{1}{M^2}E_i\partial_j\tilde{F}_{ji}
+\frac{1}{M^2}\tilde{B}^i\tilde{B}_i\nonumber\\
&&+\frac{1}{2M^2}\left(\tilde{E}^i\tilde{E}_i+\tilde{B}^i\tilde{B}_i
+E^i\nabla^2 E_i-B^i\nabla^2 B_i\right)
\label{Hcanon}
\end{eqnarray}

The first term in the \textit{r.h.s.} on the equation above is the usual energy of Maxwell fields, whilst all other terms are inversely proportional to the LW mass. However, at this point it is necessary to notice that, as widely known for the usual Abelian model\footnote{See, \textit{e.g.}, the chapter 8 on \cite{Weinberg:283223}}, from the gauge invariance, or, equivalently, from the singularity of the kinetic matrix of \eqref{LWaction}, and from the fact that the physical commutation relations between the original gauge fields and their momenta should not take the canonical form, this Hamiltonian have irrelevant degrees of freedom associated to the existence of constraints. The correct quantization method should be equivalent to the quantization in the reduced phase space, where only gauge invariant observables are quantized. Pionner work for LW model on this topic could be found in \cite{Galvao:1986yq}. The method to investigate the Hamiltonian formulation of constrained systems was developed by Dirac \cite{PSP:2025912,Dirac:1950pj,Dirac:1951zz,Dirac:1958sq}. Following the Dirac nomenclature, we have two first class (vanishing Poisson bracket) in conjugate momentum
\begin{equation}
\tilde{\Pi}^0=0\;,\qquad\Pi^0+\partial_i\tilde{\Pi}^i=0
\label{cons1}
\end{equation} 
such that this spoils the definition of Poisson brackets \eqref{PB}. Secondary constraints are associated to the necessary consistency of equations of motion for primary constraint, \textit{i.e.}, the conservation of the primary constraints in time
\begin{equation}
\partial_0\tilde{\Pi}^0=\{\tilde{\Pi}^0,H_c\}_P=0\label{cons0}
\end{equation}
\begin{eqnarray}
\partial_0\left(\Pi^0+\tilde{\Pi}^i\right)&=&\{\Pi^0+\tilde{\Pi}^i,H_c\}_P\nonumber\\
&=&\partial_i\Pi_i=0
\label{cons2}
\end{eqnarray}
All constraints, equations \eqref{cons1} and \eqref{cons2}, are first class ones, which is a consequence of the Poisson algebra of the constraints with themselves and with the canonical Hamiltonian. The dynamics of the system  is described by the total Hamiltonian, which couple these constraints via Lagrange multipliers. A new set of second class constraints emerge due to first-class constraints and they are the generators of gauge transformations. This yields the gauge ambiguity, which means that the equations of motion are still degenerate and depend on the functional arbitrariness. Besides, some Lagrange multipliers are still undetermined.  To remove this arbitrariness, one has to impose external gauge-fixing conditions for each first class constraint. For our purpose, however, the  multiplier associated to \eqref{cons0} (and his associated canonical momentum) are non-physical\footnote{Dirac quantization procedure guaranteed that physical observables are independent from these Lagrange multipliers and they do not change the counting of degrees of freedom \cite{Gitman:1624648}.}, thus, this additional part is irrelevant: in terms of the Dirac formalism, this is a first class constraint which corresponds to the (trivial) gauge symmetry  that generate arbitrary shifts in the Lagrange multiplier \cite{Henneaux:242534}. Thus, we can use the usual Dirac quantization procedure excluding this constraint. The generator of the gauge symmetry is given by
\begin{equation}
G=\int d^3x\left(A(x)\tilde{\Pi}^0+B(x)(\Pi^0+\partial_i\tilde{\Pi}^i)+C(x)\partial_i\Pi^i   \right)
\end{equation}
where $A,B,C$ are the parameters of the gauge transformation and, by the consistency equation
\begin{equation}
\partial_t G(x)=\{G(x),H_E\}\,\,\rightarrow\,\, \partial_t C=A\,\,,B=C
\end{equation}
Thus, using \eqref{cons0}
\begin{equation}
G=\int d^3x\left( \Pi^\mu\partial_\mu C(x)+\tilde{\Pi}^\mu\partial_\mu\partial_tC(x)  \right)
\end{equation}
Therefore, the gauge transformations induced by the first order constraints are given by
\begin{eqnarray}
&&\delta A_\mu =\{A_\mu,G\}=\partial_\mu C(x)\\
&&\delta \tilde{A}_\mu =\{\tilde{A}_\mu,G\}=\partial_\mu\partial_tC(x)
\end{eqnarray}
The imposition of the gauge fixing condition convert them into second class constraints and the reduction of phase space may be carried out by implementing the second class constraints in the strong sense, provided we replace all the Poisson by appropriate Dirac brackets \cite{Dirac:1951zz,Dirac:1958sq}. Since $A^0$ and $\dot{A}^0$ don't appear in the Lagrangian, we impose the temporal and the Coulomb-type gauge \cite{Fadeev:241221}
\begin{equation}
A^0=\dot{A}^0=0\,,\qquad\left(1+\frac{\partial^2}{M^2}\right)\partial_iA^i=0 
\label{cons3}
\end{equation}

A short remark: Although we are formally working with a first order theory for two fields, the full action \eqref{LWaction}, including LW-sector, is invariant under gauge transformations
\begin{equation}
A\rightarrow A^u=UAU^{-1}-\frac{i}{g}dUU^{-1}\,,\qquad U=e^{iu}
\label{AgaugeTransf}
\end{equation}
Hence, an additional fixing of the gauge for LW-parter is, apparently, unnecessary. However, is due to the imposition of a gauge condition on the HD sector too that the kinetic matrix is regularized.  
Due to the set of second class constraints, we need to introduce Dirac brackets to perform the correct quantization:
\begin{equation}
\{A(x),B(y)\}_{DB}=\{A(x),B(y)\}_P-\Delta\label{DiracBrackets}
\end{equation}
\begin{equation}
\Delta=\int d^3zd^3\bar{z}\{A(x),\phi^A(z)\}_{PB}C_{AB}^{-1}(z,\bar{z})\{\phi^B(\bar{z}),B(y)\}_P
\end{equation}
where $\phi^A$ is the set of second class constraint, equations \eqref{cons1},\eqref{cons2} and \eqref{cons3} and $C^{AB}(z,\bar{z})=\{\phi^A(z),\phi^B(\bar{z})\}$. So, defining
\begin{eqnarray}
\phi^1=\Pi^0(x)+\partial_i\tilde{\Pi}^i(x)\,,\,\,
\phi^2=\tilde{\Pi}^0(x)&&\,,\nonumber\\
\phi^3=\partial_i\Pi^i(x)\,,
\phi^4=A_0(x)&&\,,\nonumber\\
\phi^5=\tilde{A}_0(x)\,,\,\,
\phi^6=\left(1+\frac{\partial^2}{M^2}\right)\partial_iA^i(x)&&\,,
\label{constraints}
\end{eqnarray}
following \cite{Henneaux:242534} and by taking into account that the full system of constraint, \eqref{constraints}, is second class,  the number of physical degrees of freedom, $\mathcal{N}_{df}$ for this system is equal to
\begin{equation}
\mathcal{N}_{df}=\frac{16(canonical\,\,variables)-6(constraints)}{2}=5
\end{equation}
This is a very important result: the photon, as in usual Abelian theory, is transverse, i.e., has two degrees of freedom. However, the massive LW-particle has three degrees of freedom, i.e., this particle has a non-vanishing longitudinal mode, like a Proca particle. We can easily check that the only non-vanishing elements in $C_{AB}^{-1}=-C_{BA}^{-1}$ are
\begin{equation}
C_{14}^{-1}=C_{25}^{-1}=\delta^3(x-y)\,,\,\,
C_{36}^{-1}=\left(\frac{M^2}{M^2+\partial^2}\right)\frac{1}{(\partial_i^x)^2}\delta^3(x-y)
\end{equation}
Finally, by solving \eqref{DiracBrackets}, we obtain that the equal-time commutation relations for canonical variables are given by
\begin{eqnarray}
\{A_i(x),A_j(y)\}_{DB}&=&\{\tilde{A}_i(x),\tilde{A}_j(y)\}_{DB}=0\nonumber\\
\{\Pi_i(x),\Pi_j(y)\}_{DB}&=&\{\tilde{\Pi}_i(x),\tilde{\Pi}_j(y)\}_{DB}=0\nonumber\\
\{A_i(x),\Pi_j(y)\}_{DB}&=&\left(\delta_{ij}+\partial_i^x\frac{1}{\nabla^2}\partial_j^y\right)\delta^3(x-y)
\nonumber\\
\{\tilde{A}_i(x),\tilde{\Pi}_j(y)\}_{DB}&=&-\delta_{ij}\delta^3(x-y)
\label{ETCR}
\end{eqnarray}
which, by rewriting in terms of the electric and magnetic fields associated for the photon and its Lee-Wick partner is 
\begin{eqnarray}\label{conmutadorProca}
\{A_i(x),E_j(y)\}_{DB}&=&-\left(\delta_{ij}+\partial_i^x\frac{1}{\nabla^2}\partial_j^y\right)\frac{e^{-iM(x-y)}}{4\pi(x-y)}\label{comm1} \\
\{\tilde{A}_i(x),\tilde{E}_j(y)\}_{DB}&=&-M^2\delta_{ij}\delta^3(x-y)\label{comm2}
\end{eqnarray}

In the limit for LW mass going up to infinite, the Maxwell usual commutator is recovered in \eqref{comm1}; the photonic sector only feels the LW sector via the LW mass. From these results, it is important to notice that the new degrees of freedom are completely independents and, in fact, we can decouple the pure LW higher derivative sector from the photonic sector. The vacuum solution has two types of modes expansion, Proca and Maxwell-type. In the following section, we discuss the ill-behaviour for the Proca-type modes and define the first prescription to remove these degrees of freedom from the physical space.

\section{Causality violation and first prescription}\label{1stpre}

As shown in the Appendix \ref{twofieldsformulation}, LW sector can be thought of as a Proca field. In this sense, we can use a mode expansion for a massive vector field \cite{Greiner:1628401}
\begin{equation}
\tilde{A}^{\mu}=\sum_{\lambda=0}^3\int \frac{d^3p}{(2\pi)^3\sqrt{2p^0}}\left(
\tilde{a}^{\mu}_{\lambda}(\textbf{p})e^{ipx}+\tilde{a}^{\dagger\mu}_{\lambda}(\textbf{p})e^{-ipx}
\right)
\label{meLW}
\end{equation}
where $\lambda$ is a spin index, $p^0=\eta_C\omega_p=\eta_C\sqrt{\textbf{p}^2+M^2}$ is the energy of the particle and the parameter $\eta_C$ takes values $\pm 1$. Usually, the positive value is chosen to prevent instability and has a well-defined vacuum state. However, this is impossible for LW particles in the usual prescriptions. In fact, by defining the commutator signs by $\eta_C,\eta_N$
\begin{eqnarray}
[\tilde{A}_{\mu}(x),\tilde{\Pi}_{\nu}(y)]_{x^0=y^0}&=&\eta_C\delta_{\mu\nu}\delta^3(\textbf{x}-\textbf{y})
\\
\left[a(\textbf{p})_{\lambda},a_{\lambda'}^{\dagger}(\textbf{q})\right]&=&
\eta_N\delta_{\lambda\lambda '}\delta^3(\textbf{p}-\textbf{q})\label{mexp}
\end{eqnarray}
and the Hamiltonian sign $\eta_H$, then we have
\begin{equation}
\eta_C=\eta_H\eta_N \label{etas}
\end{equation}
Since the sign of the norm is defined by $\eta_C$, which must be positive by consistency and the Hamiltonian sign is $-1$ by definition of LW model, then we can only choose states of negative energy, \textit{i.e.}
\begin{equation}
\eta_C=-1\,\leftarrow\eta_H=-1\,,\quad\eta_N=1\label{etasvalues}
\end{equation}
In a relativistic classical sense, this means that LW particles are taquionic and causality is ill-defined. Another possibility is to choose negative norm states with positive energy, but we unfortunately lose the unitarity. Following the Lee prescriptions, of course, this problem can be circumvented, but with the above mentioned problems. At this point, the usual proposal to control this problem is based on the observation that in the mode expansion \eqref{meLW} the energy is always in combination with temporal coordinate. Thus, we can reabsorb the negative energy sign in time, such that LW particles are moving backward in time. In this sense, LW particles have an acausal dynamics (which means that information is sent to the past and the future dynamical evolution depend of this information), but microcausality is recovered, \textit{i.e.}, the fact that causally linked events are time-like, and the mode expansion is well-defined. To see this, using the mode expansion \eqref{meLW} and the commutation relation \eqref{mexp}, we explicitly calculate the Pauli-Jordan function, \textit{id est}, the non-equal time commutation relation for the LW field
\begin{equation}
[\tilde{A}_{\mu}(x),\tilde{A}_{\nu}(y)]
=\frac{1}{8\pi r}\frac{d}{dp}\int_{-\infty}^{+\infty}\frac{dr}{p^0}\left(
e^{i(p_0r_0+pr)}+e^{-i(p_0r_0+pr)}\right)\,,\nonumber\\
\end{equation}
where $r_0=x_0-y_0<0$, $r$ is the radial spherical coordinate and the usual orthonormal condition for the basis polarizations of massive vector field was used \cite{Greiner:1628401}. Following the Bogolyubov's prescriptions for the integration of cylidrical functions \cite{Bogolyubov:104088} and defining $\lambda=\sqrt{x^2-y^2}>0$, we can calculate this commutator in the coordinate space
\begin{equation}
[ \tilde{A}_{\mu}(x),\tilde{A}_{\nu}(y) ]=-\frac{1}{2\pi}\left(
\delta(\sqrt{\lambda})-\frac{1}{2\sqrt{\lambda}}\theta(\lambda)J_1(\sqrt{\lambda /M})\right)\label{PJ}
\end{equation}
where $\theta$ represent the Heaviside function and $J_1$ is the Bessel function of order one. For $\lambda<0$, the Pauli-Jordan function \eqref{PJ} is vanishing, then the propagation amplitude is exactly zero outside the light-cone, like it should be. In this sense, LW-partners are microcausal.

Now, we can use the Gupta-Bleuler prescription and obtain the gauge unambiguous space for this theory. Following the original idea in \cite{Gupta:1949rh,Bleuler:1950cy}, we only admitted vector states $\mid\!\!\Psi\rangle$ for which the expectation value of the gauge condition \eqref{cons3} is satisfied
\begin{equation}
\langle\Psi\!\!\mid \left(1+\frac{\partial^2}{M^2}\right)\partial_\mu A^\mu \mid\!\!\Psi\rangle =0
\label{gaugcond1}
\end{equation}
where, since in the limit to infinity for the LW mass we must recover the Maxwell theory, can perform the decomposition of the full five degrees of freedom for the potential vector as follows
\begin{eqnarray}
A_{\mu}&=&A_{\mu}^{Maxwell}+A_{\mu}^{Lee-Wick}\nonumber\\
&=&\frac{1}{(2\pi)^3}\sum_{\lambda =0}^3 \int\left\{ \frac{d^3p}{\sqrt{2p^0}}\left(a^{\mu}_{\lambda}(\textbf{p})e^{ipx}+a^{\dagger\mu}_{\lambda}(\textbf{p})e^{-ipx}\right)
+ \frac{d^3\tilde{p}}{\sqrt{2\tilde{p}^0}}\left(\tilde{a}^{\mu}_{\lambda}(\tilde{\textbf{p}})e^{i\tilde{p}x}+\tilde{a}^{\dagger\mu}_{\lambda}(\tilde{\textbf{p}})e^{-i\tilde{p} x}\right)\right\}
\end{eqnarray}
Thus, the factor $\partial^2/M^2$ just adds up a multiplicative term proportional to momentum of the particle; in the Maxwell particle case, this factor approaches zero, because the LW mass term (understood as a Pauli-Villars cut-off) is much greater than the photon momentum. Then
\begin{eqnarray}
\left(1+\frac{\partial^2}{M^2}\right)\partial_\mu A^\mu &=& \int\frac{d^3p}{(2\pi)^3}\sqrt{\frac{p^0}{2}}\left\{-i\left(a^{0}_\lambda(\textbf{p})+a^{j}_{\lambda\|}(\textbf{p})\right)e^{ipx}
+i\left(-a^{\dagger 0}_{\lambda}(\textbf{p})+a^{\dagger j}_{\lambda\|}(\textbf{p})\right))e^{-ipx}\right\}\nonumber\\
&&-i \int\frac{d^3\tilde{p}}{(2\pi)^3}\sqrt{\frac{(\tilde{p}^0)^3}{2M^2}}\left\{a^{0}_{\lambda}(\tilde{\textbf{p}})e^{i\tilde{p} x}+a^{\dagger 0}_{\lambda}(\tilde{\textbf{p}})e^{-i\tilde{p} x}\right\}
\end{eqnarray}
By applying it to \eqref{gaugcond1}, we can see that admitted states by the Gupta-Bleuler-type condition lack of temporal and longitudinal modes in photon sector, and the temporal modes in LW sector; \textit{i.e.}, these modes have a trivial norm. In this sense, the higher-derivative part in the gauge condition solely remove the non-physical temporal modes in the purely LW sector. Of course, this cannot be a correct full prescription, since the theory  is plagued by longitudinal and transverse modes of LW particles, which are anti-causal observables (ghosts) in an high enough energy regime for to be stable. Before proceeding in the analysis, we need to define the corresponding BRST operator on the which our first prescription for remove such modes will be proposed.
Considering the gauge transformation \eqref{AgaugeTransf} for the quantum operator fields $A_{\mu}$ in its infinitesimal form for Abelian case
\begin{equation}
A'_{\mu}=A_{\mu}+\varepsilon\partial_{\mu}c(x)+\mathcal{O}(\varepsilon^2)
\end{equation}
where $c(x)$ can be considered as a free quantum field. Due to the gauge condition \eqref{cons3}, this function must satisfy
\begin{equation}
\left(1+\frac{\partial^2}{M^2}\right)\partial^2c(x)=0
\end{equation}
which, defining a gauge charge operator $\mathcal{Q}$, is equivalent to
\begin{eqnarray}
[\mathcal{Q},A_{\mu}(x)]&=&\left(1+\frac{\partial^2}{M^2}\right)\partial_{\mu}c(x)\\
\mathcal{Q}&=&\int d^3x\left(1+\frac{\partial^2}{M^2}\right)\partial_{\mu}A^{\mu}
\overset{\leftrightarrow}{
\partial_0} c(x) \label{BRSTcharge}
\end{eqnarray}
This operator is the LW-Abelian version of the BRST charge operator, while $c(x)$ can be recognized as the Faddeev-Popov ghost field. Introducing the notation for the $d-transformation$, the fundamental BRST (gauge) variations
\begin{equation}
d_{\mathcal{Q}}\Psi=[\mathcal{Q},\Psi]_{\pm}
\end{equation}
where the signal in the bracket means commutator or anti-commutator, depending on whether we have bosonic or fermionic fields, respectively. Introducing the anti-ghost field $\bar{c}$ to encode the gauge condition 
\begin{equation}
d_{\mathcal{Q}}A_{\mu}=i\left(1+\frac{\partial^2}{M^2}\right)\partial_{\mu}c\,,\qquad
d_{\mathcal{Q}}c=0\,,\qquad
d_{\mathcal{Q}}\bar{c}=-i\left(1+\frac{\partial^2}{M^2}\right)\partial_{\mu}A^{\mu}\,.
\end{equation}
In the equations below, it is usual to introduce the Nakanishi-Lautrup field, a Lagrange multiplier which carries the information of the gauge condition (see, \textit{e.g.}, \cite{Nakanishi:216308}). Thus, we can rewrite the BRST transformations in the following way
\begin{eqnarray}
&d_{\mathcal{Q}}A_{\mu}=i\left(1+\frac{\partial^2}{M^2}\right)\partial_{\mu}c\,,\qquad d_{\mathcal{Q}}c=0\,,\nonumber\\
&d_{\mathcal{Q}}\bar{c}=ib^a\,,\qquad d_{\mathcal{Q}}b^a=0 \label{BRST}
\end{eqnarray}
In this notation, the Gupta-Breuler condition is equivalent to cancelling the vacuum expectation value of the Nakanishi-Lautrup field. Thus, this transformations extend the usual BRST construction for including the HD sector. With respect to this BRST-charge operator, we can define the physical subspace as the quotient of its kernel and the closure of its range \cite{Ojima:1978hy}
\begin{equation}
\mathcal{H}_{physical}=Kernel\{\mathcal{Q}\}/\overline{Ran\{\mathcal{Q}\}}
\end{equation} 
whilst the quantizing BRST invariant action for the Lee-Wick Abelian gauge theory, following the Faddev-Popov procedure \cite{Faddeev:1997aw,Faddeev:1967fc}, is given by 
\begin{eqnarray}
\mathcal{S}_{FP}&=&\int d^4\left\{-\frac{1}{4}F_{\mu\nu}F^{\mu\nu}
+\frac{1}{4M^2}(\partial_{\alpha}F_{\beta\gamma})(\partial^{\alpha}F^{\beta\gamma})\right.\nonumber\\
&&\left.+ib\left(1+\frac{\partial^2}{M^2}\right)\partial_{\mu}A^{\mu}
+\bar{c}\left(1+\frac{\partial^2}{M^2}\right)\partial^2c\right\}
\label{SFP}
\end{eqnarray}

Lastly, in order to remove of the physical space all the remaining LW modes, we can send these states to the trivial sector of the cohomology of the BRST operator by using the Nielsen-Piguet trick \cite{Nielsen:1975fs,Nielsen:1975ph,Piguet:1984js}, extending the BRST to include also the LW mass parameter, obtaining an extended Slavnov-Taylor identity that keeps control of the LW mass dependence of the gauge-invariant Green's function of the theory. Thus, as all higher derivative sector is \textit{tied up} to this mass term, then any observable can be built with them. Then, we introduce the doublet
\begin{equation}
d_{\mathcal{Q}}M^{-2}=\xi\,,\qquad d_{\mathcal{Q}}\xi=0\label{BRSText}
\end{equation} 
where $\xi$ is a Grassmann parameter with ghost number 1 \cite{Piguet:1984js,Piguet:1995er}. Observables are local composite operators belonging to the cohomology of the BRST charge operator $d_{\mathcal{Q}}$ (\textit{i.e.}, BRST/gauge invariant composite operators). (Connected) correlation functions for these operators can be obtained adding to the action \eqref{SFP} a BRST invariant term $\int dx J(x)\mathcal{O}(x)$, being $J(x)$ a BRST invariant external source, and deriving the partition function with respect to these sources
\begin{equation}
\langle \mathcal{O}(x_1)...\mathcal{O}(x_n)\rangle_{conn} =
\frac{\delta^n Z}{\delta J(x_n)...\delta J(x_1)}\vert_{J,\xi=0}\label{vevconn}
\end{equation}
Since the generating function must be invariant under BRST transformations \eqref{BRST} and \eqref{BRSText}, we have the Slanov-Taylor identity
\begin{equation}
\mathcal{B}Z=\int dx\left[(d_{\mathcal{Q}}A_{\mu}^a)\frac{\delta Z}{\delta A_{\mu}^a}  
+(d_{\mathcal{Q}}b^a)\frac{\delta Z}{\delta b^a}\right]+(d_{\mathcal{Q}}M^{-2})\frac{\delta Z}{\delta\xi}=0
\label{STI}
\end{equation}
By applying $n$ derivatives with respect to the external source, like in \eqref{vevconn}, and once in the parameter $\xi$ in the Slavnov-Taylor identity \eqref{STI}, yields the equation
\begin{equation}
\frac{\partial}{\partial M^{-2}}\left(\frac{\delta^n Z}{\delta J(x_n)...\delta J(x_1)}\right)-
\xi\frac{\partial^2}{\partial\xi\partial M^{-2}}\left(\frac{\delta^n Z}{\delta J(x_n)...\delta J(x_1)}\right)=0
\end{equation}
Afterwards, by switching off all sources and parameters, we finally get
\begin{equation}
\frac{\partial}{\partial M^{-2}}\langle \mathcal{O}(x_1)...\mathcal{O}(x_n)\rangle_{conn}=0
\end{equation}
which establishes that extending the BRST transformation for including the (inverse of the) LW mass parameter, states with the degrees of freedom of the LW sector or even the LW mass parameter, fall on the range of the BRST charge operator, \textit{ergo}, like in the case of Faddeev-Popov ghost, LW particle states are completely undetectable in the physical world (they are confined to the virtual sector). In this case, observables are the same as those in the (quantum) Maxwell case.  

\section{BRST triviality of the higher derivative term}\label{BRSTtriv}

The use of the mechanism of extended BRST \textit{\`a la Nielsen-Piguet} appears to be very artificial, but, the absence of the LW massive modes in physical space of observables is actually an effective consequence of the existence of a more general symmetry in the LW pure sector. This symmetry emerges as a natural consequence of including higher-derivative terms, and a new subsidiary condition for physical states  arises naturally: observables belong to the cohomology of the BRST and new-HD-symmetry or, in a equivalent way, observables belong to the extended super-BRST operator, id est, the cohomology operator of the graded algebra defined by of these operators, as built in \cite{Faizal:2011fp} for the topological theories.

To see this, let us consider a theory in which we switch off the photonic sector, that is, a purely higher derivative sector in \eqref{SFP}, such that the action is given by
\begin{equation}
S_{LW}^{pure}=\frac{1}{M^2}\int d^4x\left[\frac{1}{2}(\partial_\mu F^{\mu\nu})^2
-b_{\nu}\partial^{\nu}\partial_{\mu}A^{\mu}+\bar{c}_{\nu}\partial^{\nu}\partial_{\mu}c^{\mu}\right]
\label{SLWpure}
\end{equation}
where the kinetic term of field strength takes into account just the three Proca-type degrees of freedom and the Lorentzian index in the Nakanishi-Lautrup and ghost-antighost \textit{vectorial} fields come from the following redefinition
\begin{equation}
\partial_{\mu}\Psi=\Psi_{\mu}\,,\qquad \Psi=\{b,c,\bar{c}\}
\end{equation}
which correspond to completely independent degrees of freedom, as showed in the Section \ref{CQDB}, the \textit{velocity} of the auxiliary fields. This redefinition allows us rewrite the higher-derivative sector introduced by the Faddeev-Popov quantization procedure like a vectorial fields required by the BRST quartet construction, like in \cite{Rivelles:2003jd}, but in this case, they come from the ordinary Faddeev-Popov ghost associated to the Abelian LW correspondent BRST symmetry and does not spoil the nature of the Faddeev-Popov ghost term, which is rewritten in terms of massless vectorial fields which propagate because the well-defined kinetic part in \eqref{SLWpure}, but not interact, \textit{i.e.}, their behaviour is exactly equal to Faddeev-Popov ghost in photon sector. The associated Faddeev-Popov operator carries vectorial indices,  which is related to the fact that the vectorial Nakanishi-Lautrup field encoding the gauge fixing condition does not cancel the longitudinal mode of the massive gauge potential, as showed in the Section \ref{1stpre}. Now, following \cite{Rivelles:2003jd}, we can define a new \textit{higher-derivative} nilpotent BRST-type transformation, independent from the usual BRST \eqref{BRST} present in the full action, given by
\begin{equation}
\tilde{d}_{\tilde{\mathcal{Q}}}A_\mu=c_{\mu}\,,\,\,\,\tilde{d}_{\tilde{\mathcal{Q}}}c_\mu=0,\,\,\,
\tilde{d}_{\tilde{\mathcal{Q}}}\bar{c}_\mu=b_{\mu}\,,\,\,\,\tilde{d}_{\tilde{\mathcal{Q}}}b_\mu=0.
\label{HDBRST}
\end{equation}
Note that in this case the new transformation for $A_{\mu}$ is defined only on the massive degrees of freedom of the gauge vector. Notice also that unlike the \textit{genuine} BRST transformation \eqref{BRST}, this symmetry is not a true BRST symmetry because the vectorial Faddeev-Popov ghost do not define a Maurer-Cartan form like in the usual geometric interpretation of gauge potentials \cite{Piguet:1995er}. The action \eqref{SLWpure} is invariant under \ref{HDBRST}, namely is closed with respect to the $\tilde{d}_{\tilde{\mathcal{Q}}}$ operator, but don't exact, like in the topological-like models studied by Rivelles. This means that the HD action \eqref{SLWpure} can be written as
\begin{equation}
S_{LW}^{pure}=\frac{1}{M^2}\int d^4x\left[\frac{1}{2}(\partial_\mu F^{\mu\nu})^2
-\tilde{d}_{\tilde{\mathcal{Q}}}\left(\bar{c}_{\nu}\partial^{\nu}\partial_{\mu}A^{\mu}\right)\right]
\label{SLWpure2}
\end{equation}
Nevertheless, the physicality criteria of unitarity of the $S-$matrix and (BRST/HD-BRST) invariance and positivity of the states on the physical space are guaranteed \cite{Nakanishi:216308}: the LW-partner's degrees of freedom and the vectorial ghost field in \eqref{HDBRST} form a doublet\footnote{The Nakanishi-Lautrup and anti-ghost vector fields, as usual, form another doublet.}, then these fields are cohomologicaly trivial, \textit{i.e.}, belong to the trivial sector of the cohomology of the operator $\tilde{d}_{\tilde{\mathcal{Q}}}$ \footnote{See the Appendix \ref{doublets}}. Then, by using a subsidiary condition \textit{\`a la Ojima-Nakanishi}, physical states must be such that they belong to the quotient of the  kernel with the range of this charge too, in addition to the usual BRST charge defined in \eqref{BRSTcharge}, \textit{i.e.}, physical states should be identified with the quotient of the  closed (new-BRST-type invariant) states with these exact (trivial)
states, which contains the doublet fields. The independence of the observables with respect to doublet fields can be tested in the same manner as in the previous Section. This new higher derivative BRST-type symmetry is a characteristic  of the LW term, which is explicitly trivial in a cohomological sense; then, LW-parters are confined, undetectable, like Faddeev-Popov ghost particles associated to the photon sector, but, in the case in which no extending BRST extending Nielsen-Piguet's trick is implemented, observables in the theory can be functions of the mass parameter\footnote{See, \textit{e.g.}, the reference \cite{Bonin:2009je}, where the corrections to the Stefan-Boltzmann law was calculated.}.

\section{Conclusions}

The Lee-Wick model appears to be a very good alternative for particle physics, since its higher derivative terms naturally emerge as a deformation for the Abelian model whenever space-time is quantized, as predicted by quantum gravity models and extended forms of the Heisenberg's uncertainty  principle, besides solving the hierarchy  problem in the standard model for elementary particles without supersymmetry and being an effective way of regularizing quantum loops corrections, like in the Slavnov procedure. 

However, the new degrees of freedom associated to higher-derivative terms seem to exhibit a deeply exotic behaviour, which is not physically acceptable, since it violates the fundamental principles upon which a quantum theory is built up, like $S-$matrix unitary or causal propagation of particles. A set of very bizarre conditions are necessary to come over this problem, but in any case this leads to a completely satisfactory answer, in addition to the fact that ghost particles are removed by hand from the physical states.

In this paper, we propose a possible mechanism which guaranteed that new degrees of freedom are confined to the virtual sector, \textit{i.e.}, Lee-Wick model displays a ghost-free spectrum, therefore unitarity and causality are both well defined. Our prescription consists in adopting the Nielsen-Piguet trick in the usual BRST transformation to remove the dependence of observables on the Lee-Wick mass parameter, via the doublet theorem. Since all higher derivative degrees of freedom are always \textit{tied up} to this mass parameter, this mechanism removes them from the physical space. However, in a different way, this is not a new artificial prescription, but rather an effective consequence of the existence of the more fundamental BRST-type symmetry in the pure higher derivative sector, such that all the higher derivative fields belong to the trivial part of the cohomology of this new symmetry. Thus, the criterion  for defining physical states (observables) \textit{\`a la Nakanishi-Ojima} is that they are vectors belonging to the kernel of the usual BRST and the new higher derivative charge operator simultaneously but, by invoking the doublet theorem, all higher derivative degrees of freedom are excluded from observables, that is, are undetectable particles in the physical world. The crucial difference between the two prescriptions is if LW mass terms is a physical parameter or not. 

At this point, we can speculate about possible consequences of these results. First, if all higher derivative terms of Lee-Wick-type are cohomologically trivial in the sense of the first prescription, then a Slavnov regularization \cite{Bakeyev:1996is} is a natural prescription for constructing finite theories without modifying the physical content of the theory as a formal extension of Pauli-Villars regularization. Second, being the Lee-Wick terms a necessary consequence from quantizing the space-time in a quantum gravitational framework (restricted to 3+1 dimensions), then the Lee-Wick partners could represent serious candidates for the dark sector of the standard cosmology, like in the so-called \textit{quintom models} (see, \textit{e.g.}, \cite{Guo:2004fq}), such that the triviality cohomological explains the impossibility of detecting dark matter particles. The authors hope to report some results on these issues soon.

\section*{Acknowledgements}
The Coordena{\c{c}}{\~{a}}o de Aperfei{\c{c}}oamento de Pessoal de N{\'{\i}}vel Superior (CAPES) and  the Funda\c{c}\~ao Carlos Chagas Filho de Amparo \`a Pesquisa do Estado do Rio de Janeiro (FAPERJ),  are gratefully acknowledged by financial support. We express our gratitude to J. A. Helay\"el-Neto for discussions and for reading and giving suggestions on our original manuscript.
\appendix

\section{Explicit two-field formulation}
\label{twofieldsformulation}

It is very instructive to see explicitly that the LW Lagrangian \ref{LWaction} can be understood as a sum of a vector massless fields, the photon, and a vector massive fields Proca-type. To achieve this task, we note that the Lagrangian
\begin{equation}
\tilde{\mathcal{L}}_{LW}=-\frac{1}{4}F_{\mu\nu}F^{\mu\nu}-M^2\tilde{A}^2+2F^{\mu\nu}\partial_{\nu}\tilde{A}_{\nu}
\end{equation}
allows to recover the original LW Lagrangian by using the equation of motion of the field $\tilde{A}_{\mu}$, which is given by
\begin{equation}
\tilde{A}_{\mu}=-M^2\partial_{\nu}F^{\nu\mu}
\end{equation}
Then, if we perform the shift $A\rightarrow A+\tilde{A}$, whose Jacobian is unitary, we obtain a two-field decoupled lagrangian
\begin{equation}
\tilde{\mathcal{L}}_{LW}(A,\tilde{A})=-\frac{1}{4}F_{\mu\nu}F^{\mu\nu}+\frac{1}{4}\tilde{F}_{\mu\nu}
\tilde{F}^{\mu\nu}-M^2\tilde{A}_{\mu}\tilde{A}^{\mu}
\label{LWaction2}
\end{equation}

\section{Doublet theorem}
\label{doublets}

The doublet theorem  can be stated as follows \cite{Piguet:1995er}\\

$\bullet$ \textit{A set of fields $(u_i,v_i)$ which form a doublet with respect to a BRST(-type) operator $d_{\mathcal{Q}}$, \textit{i.e.}, such that $d_{\mathcal{Q}}u_i=v_i$ and $d_{\mathcal{Q}}v_i=0$, belong to the trivial sector of the cohomology of this operator.}\\

To prove this statement, we define the operator 
\begin{equation}
P=\int d^Dx\left(u_i\frac{\delta}{\delta u_i}+v_i\frac{\delta}{\delta v_i}\right)\,,\qquad
Q=\int d^Dx\,u_i\frac{\delta}{\delta v_i}
\end{equation}
where $D$ is the dimension of our space. The operator $P$ counts the number of fields of the type $(u_i,v_i)$ which are presents in a given expression. Furthermore, this operator obey the following algebra
\begin{equation}
\{d_{\mathcal{Q}},P\}=Q\,,\qquad [d_{\mathcal{Q}},P]=0\label{algrel}
\end{equation}
The cohomology of the operator $d_{\mathcal{Q}}$ is given by the solutions of the equation
\begin{equation}
d_{\mathcal{Q}}\Delta =0\label{cohoeq}
\end{equation}
Expanding the $\Delta$ function in the set of eigenfunction of the operator $P$, we have
\begin{equation}
\Delta=\sum_n\Delta_{n\geq 0}\,,\qquad P\Delta_n=n\Delta_n
\end{equation}
then \ref{cohoeq} can be rewritten in the equivalent form
\begin{equation}
d_{\mathcal{Q}}\Delta_n =0\label{cohoeqn}
\end{equation}
Finally, using the algebraic relations \ref{algrel}, we obtain
\begin{eqnarray}
\Delta &=& \Delta_0+\sum_{n\geq 1}\frac{1}{n}P\Delta_n\nonumber\\
&=&\Delta_0+\sum_{n\geq 1}\frac{1}{n}d_{\mathcal{Q}}Q\Delta_n\nonumber\\
&=&\Delta_0+d_{\mathcal{Q}}\left(\sum_{n\geq 1}\frac{1}{n}Q\Delta_n\right)
\end{eqnarray}
Hence, because the nilpotency of the BRST(-type) operator, all eigenfunctions corresponding to count a non-zero number of doublet fields are on the trivial sector of its cohomology. 



\end{document}